\documentclass{article}
\usepackage{authblk}  
\usepackage{graphicx}%
\usepackage{multirow}%
\usepackage{amsmath,amssymb,amsfonts}%
\usepackage{amsthm}%
\usepackage{mathrsfs}%
\usepackage[title]{appendix}%
\usepackage{xcolor}%
\usepackage{textcomp}%
\usepackage{manyfoot}%
\usepackage{booktabs}%
\usepackage{algorithm}%
\usepackage{natbib}
\usepackage{algorithmicx}%
\usepackage{algpseudocode}%
\usepackage{listings}%
\usepackage{subcaption}
\DeclareMathOperator*{\argmin}{arg\,min}

\begin{document}
\renewcommand{\thetable}{\arabic{table}}
\title{Wavelet Based Time Series Models with Time-Varying Thresholds}

\author[1]{Rhea Davis\thanks{Corresponding author: rheadavisc@gmail.com}}
\author[2]{N. Balakrishna\thanks{bala@iittp.ac.in; nb@cusat.ac.in}}

\affil[1]{Department of Statistics, Cochin University of Science and Technology, Kochi 682022, Kerala, India}
\affil[2]{Department of Mathematics and Statistics, Indian Institute of Technology Tirupati 517619, Andhra Pradesh, India}
\date{}  
\maketitle
\begin{abstract}
	This paper develops a threshold model with a time-varying threshold, represented using a wavelet series expansion. The model adequately captures irregular and abrupt variations, as well as smooth changes in the threshold parameter, allowing greater flexibility than Fourier-based approaches. Simulation experiments and real-data applications are used to evaluate the model's performance.
\end{abstract} 

\section{Introduction}\label{sec1}
Linear time series models, especially those developed under the Box–Jenkins methodology, have historically dominated time series analysis. However, these models commonly do not capture the asymmetries as well as other nonlinearities observed in many real-world time series. In such cases, piecewise linear models provide a more effective alternative (\citeauthor{tsay2005analysis}, \citeyear{tsay2005analysis}). A classical and useful model in this class is the threshold autoregressive (TAR) model introduced by \citeauthor{tong1978threshold} (\citeyear{tong1978threshold}). The TAR model is a piecewise locally linear model in which the parameters are regime-dependent, and “locally” refers to regions in the state space rather than to specific time points (\citeauthor{yadav1994threshold}, \citeyear{yadav1994threshold}). Specifically, the self-exciting TAR (SETAR) model uses lagged variables to define the threshold. A two-regime SETAR(1) model is defined as follows (\citeauthor{tsay2005analysis}, \citeyear{tsay2005analysis}): 

\begin{equation}\label{constthreshmod}
	y(t) = \begin{cases}
		\phi_0^{(1)} + \phi_1^{(1)}y(t-1) + \epsilon(t),\,\, y(t-1) \leq \gamma \\
		\phi_0^{(2)} + \phi_1^{(2)}y(t-1) + \epsilon(t),\,\, y(t-1) > \gamma
	\end{cases},\,\, t = 0, 1, \dotsc, 
\end{equation}
where $\{\epsilon_t\}$ is an independent and identically distributed (iid) sequence of mean 0 and variance $\sigma^2$ random variables, which will hereafter be denoted by $\epsilon(t) \sim \text{iid}\,(0,\sigma^2)$. For ergodicity, it is assumed that $\phi_1^{(1)}<1, \phi_1^{(2)}<1, \phi_1^{(1)}\phi_1^{(2)}<1$  (see \citeauthor{petruccelli1984threshold}, \citeyear{petruccelli1984threshold}). Here, $\gamma$ is the threshold parameter. TAR models can represent a variety of nonlinear phenomena, such as limit cycles, amplitude-dependent frequencies, and jump phenomena (\citeauthor{tsay1989testing}, \citeyear{tsay1989testing}). As a result, they are widely applied in statistics, especially in financial time series, which usually display strong signs of nonlinearity, partly due to the volatility of the global economy and the liberalization of capital flows (\citeauthor{sirikanchanarak2016time}, \citeyear{sirikanchanarak2016time}). Examples include modeling unemployment
 (\citeauthor{montgomery1998forecasting}, \citeyear{montgomery1998forecasting}), GNP (\citeauthor{li2012least}, \citeyear{li2012least}), stock prices (\citeauthor{narayan2006behaviour}, \citeyear{narayan2006behaviour}), exchange rates (\citeauthor{clements1999monte}, \citeyear{clements1999monte}), power system load (\citeauthor{huang1997short}, \citeyear{huang1997short}), and epidemiological time series (\citeauthor{watier1995modelling}, \citeyear{watier1995modelling}). For further reading, see \citeauthor{hansen1997inference} (\citeyear{hansen1997inference}) for inference and \citeauthor{hansen2011threshold} (\citeyear{hansen2011threshold}) for economic applications. 

In standard threshold autoregressive modeling, the threshold parameter is typically assumed to be time-invariant. However, this time-homogeneity assumption is often unrealistic as economic and financial systems rarely have parameters that stay fixed (\citeauthor{zhu2017asymmetry}, \citeyear{zhu2017asymmetry} and \citeauthor{yang2021threshold}, \citeyear{yang2021threshold}). This motivates the development of threshold models with time-varying thresholds. The key reasoning is that whether an economic variable is “high” or “low” is typically relative to other changing economic variables (\citeauthor{dueker2013state}, \citeyear{dueker2013state}). Ignoring the time-varying nature of thresholds can lead to model misspecification, biased estimation, and unreliable conclusions (\citeauthor{zhu2017asymmetry}, \citeyear{zhu2017asymmetry}). Recent works have considered non-constant threshold parameters, for example, see \citeauthor{bessec2003asymmetric} (\citeyear{bessec2003asymmetric}), \citeauthor{zhu2017asymmetry} (\citeyear{zhu2017asymmetry}), and \citeauthor{yu2021threshold} (\citeyear{yu2021threshold}). Prominent among them is \citeauthor{yang2021threshold} (\citeyear{yang2021threshold}), which incorporates a time-varying threshold estimated by Fourier approximation, given by 
\begin{equation}\label{tvarhreshmod}
	y(t) = \begin{cases}
		\phi_0^{(1)} + \phi_1^{(1)}y(t-1) + \epsilon(t), \quad \quad y(t-1) \leq \gamma (\frac{t}{T}) \\
		\phi_0^{(2)} + \phi_1^{(2)}y(t-1) + \epsilon(t), \quad \quad y(t-1) > \gamma(\frac{t}{T})
	\end{cases}, t = 0, 1, \dotsc, T-1,
\end{equation}
where $\epsilon(t) \sim \text{iid}\,(0,\sigma^2)$ ,  $\phi_1^{(1)}<1, \phi_1^{(2)}<1, \phi_1^{(1)}\phi_1^{(2)}<1$, and $\gamma(\cdot)$ is the time-varying threshold parameter, where it is assumed to have the Fourier approximation
  \begin{equation}\label{fouapp}
	\gamma \left(\frac{t}{T}\right) = \gamma_0 + \gamma_1\text{sin}\left(\frac{2\pi kt}{T}\right)  + \gamma_2\text{cos}\left(\frac{2\pi kt}{T}\right), t = 0, \dotsc, T-1,
\end{equation}
where $T$ is the sample size and $k$ is to be determined.

Fourier approximation is most effective when the threshold changes gradually, but less so when it changes abruptly. In many situations, threshold values may change suddenly, for example, because of market interventions or rapid policy transitions. Thus, there is a need for models that can accommodate irregular or abrupt variations in the threshold parameter.

The rest of the article is arranged as follows. Section 2 presents a concise introduction to wavelet analysis. The proposed wavelet-based threshold model is described in Section 3. Section 4 contains the simulation studies of our proposed model. The bootstrap confidence intervals for the wavelet-based threshold model are presented in Section 5. Section 6 provides a real-data application demonstrating the effectiveness of the proposed model. The conclusion of this study is given in Section 7. 

\section{Wavelet Analysis}
Wavelets have become a prominent and useful tool for nonparametric curve estimation. A major advantage of wavelet methods is their capability to adapt to irregular features, such as abrupt changes, discontinuities, and varying degrees of smoothness in unknown functions. Wavelets provide simultaneous time–frequency localization, making them useful for studying nonstationary data. For foundational information on wavelets, see \citeauthor{daubechies1992ten} (\citeyear{daubechies1992ten}). Some recent works of wavelet applications in statistics can be found in \citeauthor{bak2025adaptive} (\citeyear{bak2025adaptive}) and \citeauthor{chen2025time} (\citeyear{chen2025time}).

Let $L^2(\mathbb{R})$ denote the space of functions that are square–integrable. Then, any function $f \in L^2(\mathbb{R})$ admits an expansion in terms of dilations and translations of two fundamental functions: the father wavelet $\varphi(\cdot)$ and the mother wavelet $\psi(\cdot)$. These fundamental functions can be defined via the two–scale equations
\begin{equation} 
	\varphi(t) = \sqrt{2}\sum_k \ell_k \varphi(2t-k),
	\qquad
	\psi(t) = \sqrt{2}\sum_k h_k \varphi(2t-k).
\end{equation} 
Here, $\ell_k$ and $h_k$ are the low-pass and high-pass filter coefficients, respectively, where $\sum_{k=0}^{L-1} \ell_k = \sqrt{2}$, $\sum_{k=0}^{L-1} h_k = 0$, $h_k = (-1)^k \overline{\ell_{L-1-k}}$, and $L$ is the filter length. The translated and dilated versions of father and mother wavelets are $\varphi_{j,k}(t) = 2^{j/2}\varphi(2^jt-k)$ and $\psi_{j,k}(t) = 2^{j/2}\psi(2^jt-k)$, respectively. 

In this work, the Daubechies extremal phase family, $\text{D}(N)$, and the Daubechies least asymmetric family, $\text{LA}(N)$, are considered, where $N$ indicates the number of vanishing moments. In both families, the father and mother wavelets have compact support, allowing them to capture local features of a signal. Specifically, the support of $\varphi$ is $[0,2N-1]$, and that of $\psi$ is $[-N+1,N]$. The values of the father and mother wavelets can be computed using the Daubechies–Lagarias algorithm (see \citeauthor{morettin2017wavelets}, \citeyear{morettin2017wavelets}). The filter coefficients of these families are available in many sources, for example, see \citeauthor{daubechies1992ten} (\citeyear{daubechies1992ten}).

A signal can be studied at multiple levels using multiresolution analysis (MRA) offered by wavelets. Coarse resolutions capture the signal's low-frequency components and primary features, whereas finer resolutions preserve its high-frequency components. An MRA consists of an increasing sequence of closed subspaces $V_j$, $j\in\mathbb{Z}$ with the following properties (\citeauthor{daubechies1992ten}, \citeyear{daubechies1992ten}):

\begin{enumerate}
	\item $\displaystyle \bigcap_j V_j = \{0\}$
	\item $\displaystyle L^2(\mathbb{R}) = \overline{\bigcup_j V_j}$
	\item There exists a scaling function $\varphi \in V_0$ such that 
	$\{\varphi(\cdot-k), k\in\mathbb{Z}\}$ is an orthonormal basis for $V_0$.
	\item For all $k\in\mathbb{Z}$, 
	$g(\cdot)\in V_0 \Leftrightarrow g(\cdot-k)\in V_0$
	\item $g(\cdot)\in V_j \Leftrightarrow g(2\cdot)\in V_{j+1}$. 
\end{enumerate}

The collection $\{\varphi_{0,k}(t), \psi_{j,k}(t): j\geq 0, k\in\mathbb{Z}\}$ forms an orthonormal basis for $L^2(\mathbb{R})$. Therefore, any $f(t)\in L^2(\mathbb{R})$ can be represented as
\begin{equation}\label{wavseriesrep} 
	f(t) = \sum_k c_{0,k}\varphi_{0,k}(t) 
	+ \sum_{j\geq 0}\sum_k d_{j,k}\psi_{j,k}(t),
\end{equation} 
whose coefficients are defined by 
\begin{equation} 
	c_{0,k} = \int_{-\infty}^{\infty} f(t)\varphi_{0,k}(t)\,dt,
	\qquad
	d_{j,k} = \int_{-\infty}^{\infty} f(t)\psi_{j,k}(t)\,dt.
\end{equation}  
 The first component of \eqref{wavseriesrep} describes the coarse resolution (low-frequency) information of the function at level (0). The second component represents the remaining details, where the collection $\{\psi_j,k\}$ captures finer details of the function successively as the resolution level increases. In 
real-world applications, if higher-resolution components contribute minimally, 
the function can be suitably represented by truncating the wavelet series at a 
finite resolution level $J$. This truncated expansion corresponds to projecting 
$f(t)$ onto the subspace $V_J$. The truncated wavelet series expansion of $f$ is

\begin{equation} 
	f(t) = \sum_k c_{0,k}\varphi_{0,k}(t) 
	+ \sum_{j=0}^{J-1}\sum_k d_{j,k}\psi_{j,k}(t).
\end{equation} 
The simplest wavelet basis for $L^2(\mathbb{R})$ is the Haar basis, whose father and mother wavelets are (\citeauthor{haar1910theorie}, \citeyear{haar1910theorie})
\begin{equation}\label{haarfath}
	\varphi(t)=
	\begin{cases}
		1, & 0\leq t < 1 \\
		0, & \text{otherwise}
	\end{cases}\,\, \text{and} 
\end{equation}

\begin{equation}\label{haarmoth}
	\psi(t)=
	\begin{cases}
		1, & 0\leq t < \tfrac{1}{2} \\
		-1, & \tfrac{1}{2}\leq t < 1 \\
		0, & \text{otherwise}
	\end{cases},
\end{equation}
respectively. 

\section{Wavelet-Based Time-Varying Threshold Model}
The wavelet-based threshold model, where the threshold parameter varies over time, is described as
\begin{equation}\label{wavethreshmod}
	y(t) = \begin{cases}
		\phi_0^{(1)} + \phi_1^{(1)}y(t-1) + \epsilon(t), \,\,y(t-1) \leq \gamma (\frac{t}{T}) \\
		\phi_0^{(2)} + \phi_1^{(2)}y(t-1) + \epsilon(t),\,\,y(t-1) > \gamma(\frac{t}{T})
	\end{cases},\,\, t = 0, \dotsc, T-1,
\end{equation}
where $\epsilon(t) \sim \text{iid}\,(0,\sigma^2)$, $\phi_1^{(1)}<1, \phi_1^{(2)}<1, \phi_1^{(1)}\phi_1^{(2)}<1$, and $\gamma(\cdot)$, whose domain is [0,1), is approximated by the wavelet series expansion
\begin{equation}\label{wavserexp}
	\gamma_{\boldsymbol{\theta}} \left(\frac{t}{T}\right) = c_{00}\varphi_{00}\left(\frac{t}{T}\right) + \sum_{j=0}^{J-1}\sum_{k=0}^{2^j-1} d_{jk}\psi_{jk}\left(\frac{t}{T}\right),\, \, t = 0, \dotsc, T-1,
\end{equation} 
where $J$ is the resolution level chosen. Often, lower values of J are sufficient. Also, $\boldsymbol{\theta} = (c_{00}; d_{jk}, j=0,\dotsc,J-1, k=0,\dotsc,2^j-1)^{\top}$, where $\top$ denotes transpose. The Haar wavelet, by virtue of its definition, is naturally suited for function estimation on the unit interval. However, other wavelets suffer from boundary effects, which can be minimized by adopting the mirroring method, as suggested by \citeauthor{hardle2012wavelets} (\citeyear{hardle2012wavelets}). In this method, estimation is performed by extending the data on both ends by reversing it and appending it to both sides. Finally, the resulting estimator is truncated back to the original [0,1] interval. This approach preserves the continuity of the function, yielding a curve that is much smoother and more reliable at the dataset's extremes. The parameters to be estimated are $\boldsymbol{\theta} \in \mathbb{R}^{2^J}$, which consists of the wavelet series coefficients, as well as the autoregressive parameters $\phi_0^{(1)}$, $\phi_1^{(1)}$, $\phi_0^{(2)}$, $\phi_1^{(2)}$, and $\sigma^2$. The estimation procedure is similar to the one in \citeauthor{yang2021threshold} (\citeyear{yang2021threshold}), which is briefly described below. 
\begin{equation} \label{betaexp}
\text{Let} \,\,	\boldsymbol{\beta} = \begin{pmatrix}
		\phi_0^{(2)} & \phi_1^{(2)} & \phi_0^{(1)}-\phi_0^{(2)} & \phi_1^{(1)}-\phi_1^{(2)}
	\end{pmatrix}^{\top} = \begin{pmatrix}
		\beta_1 & \beta_2 & \beta_3 & \beta_4
	\end{pmatrix}^{\top},
\end{equation} 
and
\begin{equation}
	\textbf{x}_{\boldsymbol{\theta}}(t) = \begin{pmatrix}
		1 \\
		y(t-1) \\
		I\left(y(t-1)\leq \gamma_{\boldsymbol{\theta}}\left(\frac{t}{T}\right)\right) \\
		y(t-1)
		I\left(y(t-1)\leq \gamma_{\boldsymbol{\theta}}\left(\frac{t}{T}\right)\right)
	\end{pmatrix}, \, t = 0, \dotsc, T-1,
\end{equation}
where $I(\cdot)$ is the indicator function. Then, \eqref{wavethreshmod} becomes
\begin{equation}
	y(t) = \boldsymbol{\beta}^{\top}\textbf{x}_{\boldsymbol{\theta}}(t) + \epsilon(t).
\end{equation}
This reparameterization enables the model to be written as a linear regression conditional on $\boldsymbol{\theta}$ and hence the estimation can be performed via conditional least squares. Assume $\boldsymbol{\theta} \in \Gamma$, where $\Gamma$ is assumed to be a compact subset of $\mathbb{R}^{2^J}$. The estimation of $\boldsymbol{\theta}$ is performed by
\begin{equation}\label{thetaestexp}
	\hat{\boldsymbol{\theta}} = \argmin\limits_{\boldsymbol{\theta}\in \Gamma} \text{SSR}_{\boldsymbol{\theta}},
\end{equation}
where 
\begin{align}
	\text{SSR}_{\boldsymbol{\theta}} &= \sum_{t=1}^{T-1} \left( y(t) - \hat{\boldsymbol{\beta}}_{\boldsymbol{\theta}}^{\top} \mathbf{x}_{\boldsymbol{\theta}}(t) \right)^2 \quad \text{and}\\
	\hat{\boldsymbol{\beta}}_{\boldsymbol{\theta}} &= \left[ \sum\limits_{t=1}^{T-1} \mathbf{x}_{\boldsymbol{\theta}}(t) \mathbf{x}_{\boldsymbol{\theta}}^{\top}(t)\right]^{-1}\left[\sum\limits_{t=1}^{T-1} \mathbf{x}_{\boldsymbol{\theta}}(t) y(t)\right].
\end{align}
Note that for each $\boldsymbol{\theta}$, $\boldsymbol{\beta}$ is estimated by conditional least squares. The estimates were obtained in R with the help of the optim function, with initial values determined by the DEoptim function, which is available in the \texttt{DEoptim} package (see \citeauthor{deoptimpack}, \citeyear{deoptimpack}). Once $\hat{\boldsymbol{\theta}}$ is obtained, $\hat{\boldsymbol{\beta}}$ is determined by
\begin{equation}\label{betaest}
	\hat{\boldsymbol{\beta}}_{\hat{\boldsymbol{\theta}}} = \left[ \sum\limits_{t=1}^{T-1} \mathbf{x}_{\hat{\boldsymbol{\theta}}}(t) \mathbf{x}_{\hat{\boldsymbol{\theta}}}^{\top}(t)\right]^{-1}\left[\sum\limits_{t=1}^{T-1} \mathbf{x}_{\hat{\boldsymbol{\theta}}}(t) y(t)\right] = \begin{pmatrix}
		\hat{\beta}_1 & \hat{\beta}_2 & \hat{\beta}_3 & \hat{\beta}_4
	\end{pmatrix}^{\top}.
\end{equation}
Using \eqref{betaexp} and \eqref{betaest}, we get, 
\begin{equation}\label{arestexp}
	\hat{\phi}_0^{(1)} = \hat{\beta}_1 + \hat{\beta}_3, \, \hat{\phi}_1^{(1)} = \hat{\beta}_2 + \hat{\beta}_4, \, \hat{\phi}_0^{(2)} = \hat{\beta}_1, \text{and} \, \, \hat{\phi}_1^{(2)} = \hat{\beta}_2.
\end{equation}
The variance $\sigma^2$ is estimated as
\begin{equation}\label{varestexp}
	\hat{\sigma}^2 = \frac{1}{T-1}\sum\limits_{t=1}^{T-1} \hat{\epsilon}_t^2,
\end{equation}
where 
\begin{equation}\label{resexp}
	\hat{\epsilon}_t = y(t)-\hat{\boldsymbol{\beta}}_{\hat{\boldsymbol{\theta}}}^{\top}\textbf{x}_{\hat{\boldsymbol{\theta}}}(t)
\end{equation}
is the $\text{t}^{\text{th}}$ residual. Finally, the estimated time-varying threshold is obtained by
\begin{equation}\label{gammaestexp}
	\hat{\gamma}_{\hat{\boldsymbol{\theta}}}\left(\frac{t}{T}\right) = \hat{c}_{00}\varphi_{00}\left(\frac{t}{T}\right) + \sum_{j=0}^{J-1}\sum_{k=0}^{2^j-1} \hat{d}_{jk}\psi_{jk}\left(\frac{t}{T}\right),\, \, t = 0, \dotsc, T-1.
\end{equation}

\section{Simulation Studies}
This section contains two simulation experiments to illustrate the concepts developed in the preceding sections.
\subsection{Simulation Study - 1}
The model considered is 
\begin{equation}
	y(t) = \begin{cases}
		0.5-0.3y(t-1)+\epsilon(t), \, y(t-1) \leq \gamma \left(\frac{t}{T}\right)\\
		1 + 0.3y(t-1)+\epsilon(t), \, y(t-1) > \gamma \left(\frac{t}{T}\right)
	\end{cases}, \,\, t = 0, \dotsc, 2047,
\end{equation}
where 
\begin{equation}\label{sim1func}
	\gamma \left(\frac{t}{T}\right) = \begin{cases}
		1, \quad \frac{t}{T}<0.25\\
		1.5, \quad 0.25 \leq \frac{t}{T} <0.75 \\
		1, \quad \frac{t}{T}\geq 0.75
			\end{cases},  \,\, t = 0, \dotsc, 2047,
\end{equation}
and $\epsilon(t) \sim N(0,\sigma^2)$, where $\sigma^2 = 2.$ The step function in \eqref{sim1func} is considered to evaluate how the proposed model performs when the underlying threshold changes abruptly over time. The Haar wavelet, with its corresponding father and mother wavelets provided in \eqref{haarfath} and \eqref{haarmoth}, respectively, was used for the analysis. The resolution level $J=2$ was selected based on the nature of the time-varying threshold function considered. For each replication, $T=2048$ observations were generated from the model, and the study was conducted over 500 replications. The autoregressive estimates and corresponding root mean square errors (RMSEs) are given in Table~\ref{simres}. The true and estimated threshold function are displayed in Figure \ref{figsim1}. The results from the simulations indicate good model performance despite abrupt changes in the threshold function.
\subsection{Simulation Study - 2}
\begin{figure}
	\centering
	
	\begin{subfigure}[t]{0.45\textwidth}
		\centering
		\includegraphics[width=\textwidth]{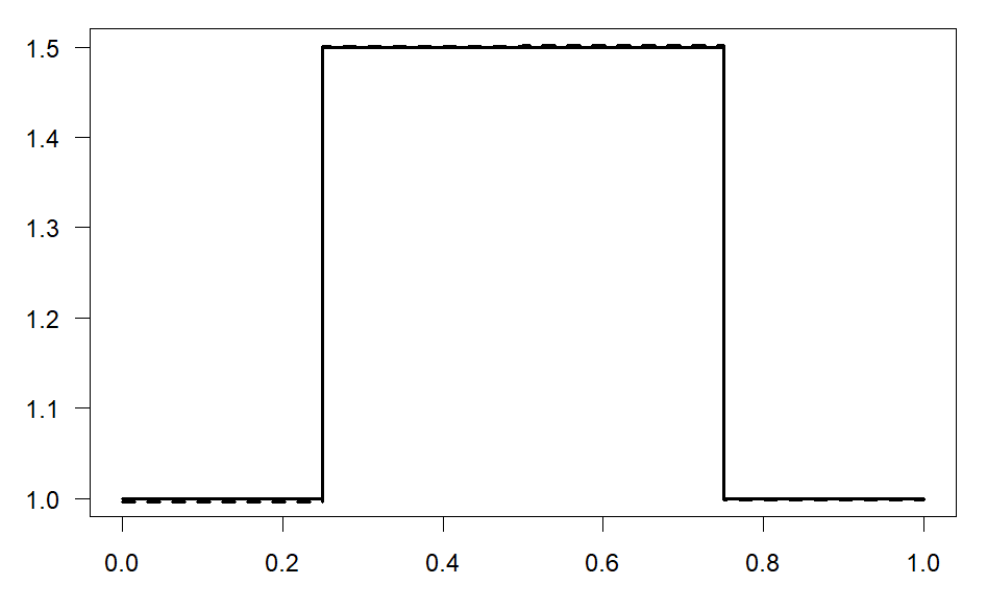}
		\caption{}
		\label{figsim1}
	\end{subfigure}
	\hfill
	\begin{subfigure}[t]{0.45\textwidth}
		\centering
		\includegraphics[width=\textwidth]{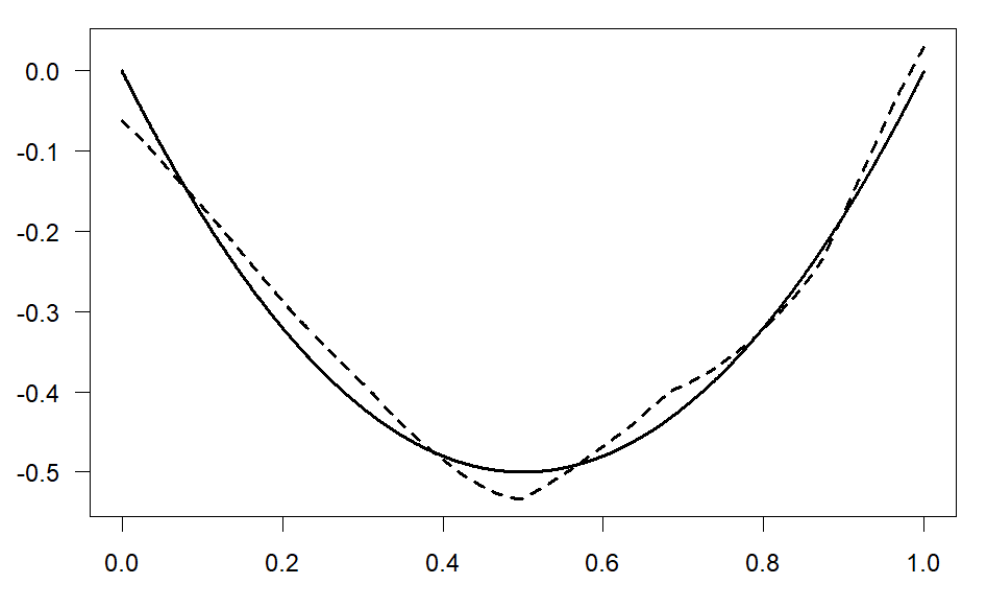}
		\caption{}
		\label{figsim2}
	\end{subfigure}
	
	\caption{True (solid) and estimated (dotted) threshold functions corresponding to Simulation Studies 1 and 2 (panels (a) and (b), respectively).}
	\label{figsim}
	
\end{figure}

For this study, the model is described as
\begin{equation}
	y(t) = \begin{cases}
		0.5+0.3y(t-1)+\epsilon(t), \, y(t-1) \leq \gamma \left(\frac{t}{T}\right)\\
		-1.0 + 0.5y(t-1)+\epsilon(t), \, y(t-1) > \gamma \left(\frac{t}{T}\right)
	\end{cases}, t = 0, \dotsc, 2047
\end{equation}
where 
\begin{equation}\label{sim2func}
	\gamma \left(\frac{t}{T}\right) = -0.5 + 0.5\left(\frac{2t}{T}-1\right)^2, t = 0, \dotsc, 2047
\end{equation}
and $\epsilon(t) \sim N(0,\sigma^2)$, where $\sigma^2 = 1.$ To study the behaviour of the proposed wavelet-based threshold model when the threshold evolves smoothly over time, the function specified in \eqref{sim2func} is considered. The wavelet employed is the Daubechies least asymmetric wavelet with four vanishing moments (LA(4)). The value of J from 2, 3, 4, and 5 that yields the lowest root mean squared error was chosen. For each replication, a sample of size $T=2048$ was generated from the proposed model, and the entire experiment was repeated 500 times. Table~\ref{simres} reports the estimates of the autoregressive parameters, together with their RMSEs. A comparison between the true threshold function and its estimate obtained from the proposed method is presented in Figure \ref{figsim2}. 
The optimal value of $J$ is 3. The results show that the proposed model provides reliable estimates and continues to perform satisfactorily even when the underlying threshold function varies smoothly.
\begin{table}	
	\caption{Results of Simulation Study - 1 and Simulation Study - 2.}
	\label{simres}
	\begin{tabular}{cccccccc}
		\toprule
		\textbf{}          & \multicolumn{3}{c}{\textbf{Simulation Study - 1}} & \textbf{} & \multicolumn{3}{c}{\textbf{Simulation Study - 2}} \\
		\textbf{Parameter} & \textbf{True} & \textbf{Estimate} & \textbf{RMSE} & \textbf{} & \textbf{True} & \textbf{Estimate} & \textbf{RMSE} \\
		$\phi_0^{(1)}$     & 0.5           & 0.4926            & 0.0429        &           & 0.5           & 0.4618            & 0.0842       \\
		$\phi_1^{(1)}$     & -0.3          & -0.3089           & 0.0448        &           & 0.3           & 0.2776            & 0.0560       \\
		$\phi_0^{(2)}$     & 1.0           & 1.0298            & 0.1285        &           & -1.0          & -0.9923           & 0.0421        \\
		$\phi_1^{(2)}$     & 0.3           & 0.2904            & 0.0483        &           & 0.5           & 0.4898            & 0.0501        \\
		$\sigma^2$         & 2.0           & 1.9938            & 0.0642        &           & 1.0           & 1.0116            & 0.0334   \\  
		\bottomrule  
	\end{tabular}
\end{table}
\section{Bootstrap Inference for Wavelet-Based Threshold Model}\label{bootsec}
This section presents a bootstrap approach to constructing confidence intervals for the wavelet-based threshold model. The algorithm for constructing the intervals is presented below.
\begin{enumerate}
	\item Generate data $y(0), \dotsc y(T-1)$ from \eqref{wavethreshmod}.
	\item Compute $\hat{\phi}_0^{(1)}, \hat{\phi}_1^{(1)}, \hat{\phi}_0^{(2)}, \hat{\phi}_1^{(2)}, \hat{\sigma}^2$ and $\hat{\gamma}_{\hat{\boldsymbol{\theta}}}(\cdot)$ as given in \eqref{arestexp}, \eqref{varestexp}, and \eqref{gammaestexp}.
	\item Obtain the residuals using \eqref{resexp} and center them by subtracting the mean. Let it be denoted by $\overset{*}{\epsilon}(t), t = 1, \dotsc, T-1.$
	\item Generate bootstrap residuals $\{\overset{*}{\epsilon}_b(t)\}, t = 1, \dotsc, T-1$, by sampling with replacement from $\{\overset{*}{\epsilon}(t)\}, t = 1, \dotsc, T-1$. 
	\item Simulate bootstrap observations $y_b(1), \dotsc y_b(T-1)$ using 
	\begin{equation*}
		y_b(t) = \begin{cases}
			\hat{\phi}_0^{(1)} + \hat{\phi}_1^{(1)}y_b(t-1)+\overset{*}{\epsilon}_b(t), \quad \quad y_b(t-1) \leq \hat{\gamma}_{\hat{\boldsymbol{\theta}}}\left(\frac{t}{T}\right)\\
			\hat{\phi}_0^{(2)} + \hat{\phi}_1^{(2)}y_b(t-1)+\overset{*}{\epsilon}_b(t), \quad \quad y_b(t-1) > \hat{\gamma}_{\hat{\boldsymbol{\theta}}}\left(\frac{t}{T}\right)
		\end{cases}, t = 1, \dotsc, T-1.
	\end{equation*} 
	Here, the starting value $y_b(0)$ is chosen to be $y(0)$. 
	\item Compute $\phi_{0,b}^{(1)}, \phi_{1,b}^{(1)}, \phi_{0,b}^{(2)}, \phi_{1,b}^{(2)}, \sigma_b^2$ and $\gamma_{\hat{\boldsymbol{\theta}},b}(\cdot)$ using $y_b(0), \dotsc, y_b(T-1)$.
	\item Steps 4-6 are repeated for a total of $B$ iterations.
	\item Suppose the $B$ bootstrap estimates of $\phi_0^{(1)}$ are given by 
	$\left\{\phi_{0,b}^{(1)} : b = 1, \dots, B \right\}$. 
	Let $F(g) = P\left(\phi_0^{(1)} \leq g \right)$ denote the cumulative distribution function of $\phi_0^{(1)}$. 
	An empirical approximation of $F(g)$ based on $B$ bootstrap replicates is given by
	\begin{equation*}
		F_B(g) = \frac{\text{Number of } \phi_{0,b}^{(1)} \le g \text{ in } \left\{ \phi_{0,b}^{(1)} : b = 1, \dots, B \right\}}{B}.
	\end{equation*}  
	Then, a $100\alpha\%$ bootstrap confidence interval for $\phi_0^{(1)}$ is $[Q_B(0.5-0.5\alpha),Q_B(0.5+0.5\alpha)]$, where $Q_B = F_B^{-1}$ is the empirical quantile function. In the same way, bootstrap confidence intervals for the other autoregressive parameters are constructed. 
	\item In the case of time-varying threshold parameter $\gamma_{\boldsymbol{\theta}}(\cdot)$, we consider the simultaneous bands via sup-t (see \citeauthor{montiel2019simultaneous}, \citeyear{montiel2019simultaneous}) described as follows:
	\begin{enumerate}
		\item For each bootstrap replication, $b = 1, \dotsc, B$, compute
		\begin{equation*}
			M_b = \sup_{t}
			\frac{\left|\gamma_{\hat{\boldsymbol{\theta}},b}\left(\frac{t}{T}\right)
				-\bar{\gamma}_{\hat{\boldsymbol{\theta}},b}\left(\frac{t}{T}\right)\right|}
			{\operatorname{sd}\!\left(\gamma_{\hat{\boldsymbol{\theta}}, b}\left(\frac{t}{T}\right)\right)},
		\end{equation*}
		where $\bar{\gamma}_{\hat{\boldsymbol{\theta}},b}(\cdot) = \frac{1}{B}\sum \limits_{b=1}^B \gamma_{\hat{\boldsymbol{\theta}}, b}(\cdot)$ and  $\operatorname{sd}\left(\gamma_{\hat{\boldsymbol{\theta}}, b}(\cdot)\right)$ denotes the standard deviation taken across bootstrap replications $b=1,…,B$.
		\item Let $c_{0.95}$ be the 95th percentile of $\{M_1, \dotsc, M_B\}$.
		\item The simultaneous confidence band is 
		\begin{equation*}
			\left[
			\hat{\gamma}_{\hat{\boldsymbol{\theta}}}\left(\frac{t}{T}\right)
			- c_{0.95}\operatorname{sd}\left(\gamma_{\hat{\boldsymbol{\theta}},b}\left(\frac{t}{T}\right)\right),\ 
			\hat{\gamma}_{\hat{\boldsymbol{\theta}}}\left(\frac{t}{T}\right)
			+ c_{0.95}\operatorname{sd}\left(\gamma_{\hat{\boldsymbol{\theta}},b}\left(\frac{t}{T}\right)\right)
			\right].
		\end{equation*}
	\end{enumerate}
\end{enumerate}
Here, we consider the same model used in Simulation Study - 1. A total of $B =1000$ bootstrap replications were used. The coverage probabilities were computed by performing the procedure 1000 times. The coverage probabilities of $\phi_0^{(1)}, \phi_1^{(1)}, \phi_0^{(2)}, \phi_1^{(2)}$ and $\sigma^2$ obtained are 0.94, 0.95, 0.93, 0.94, and 0.92, respectively. The coverage probability of $\gamma_{\boldsymbol{\theta}}(\cdot)$ based on simultaneous bands via sup-t is 0.94.  
\section{Data Analysis}
\begin{figure}
	\centering
	\begin{subfigure}[t]{0.48\textwidth}
		\centering
		\includegraphics[width=\textwidth]{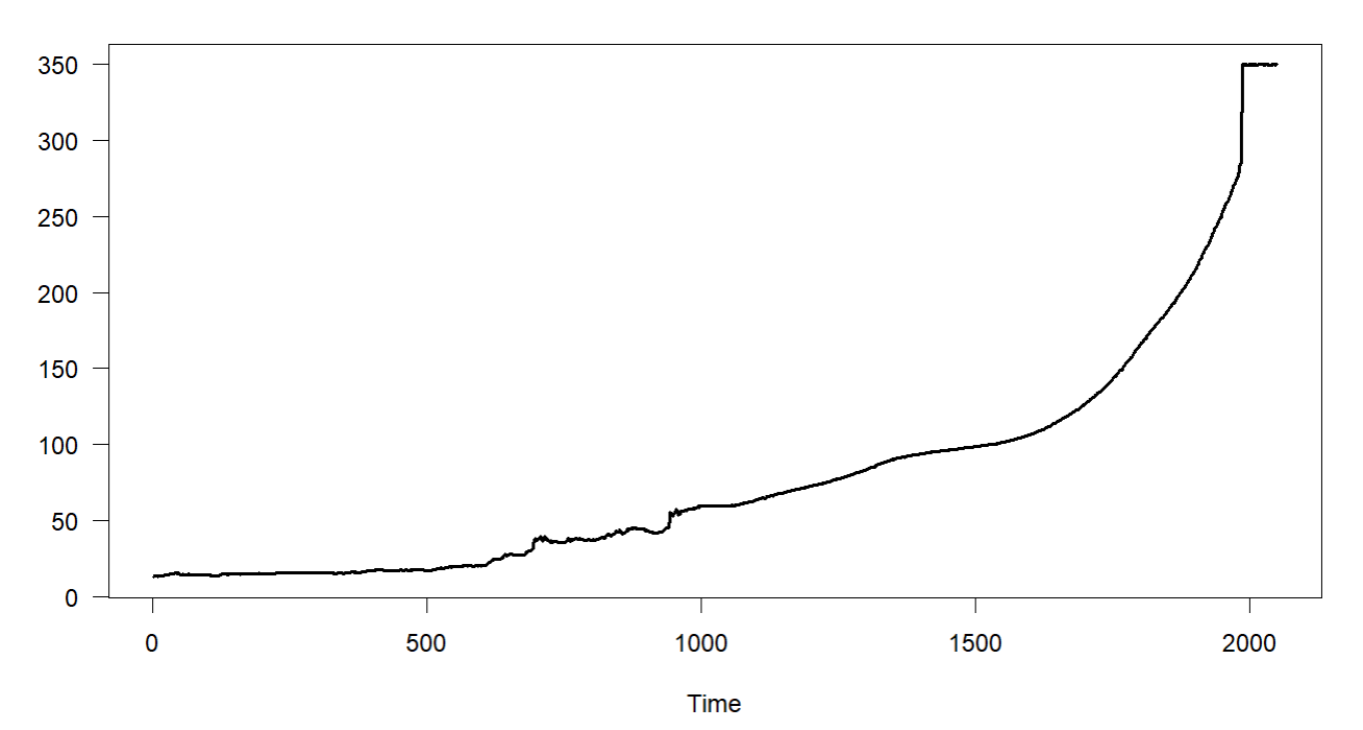}
		\caption{}
	\end{subfigure}
	\hfill
	\begin{subfigure}[t]{0.51\textwidth}
		\centering
		\includegraphics[width=\textwidth]{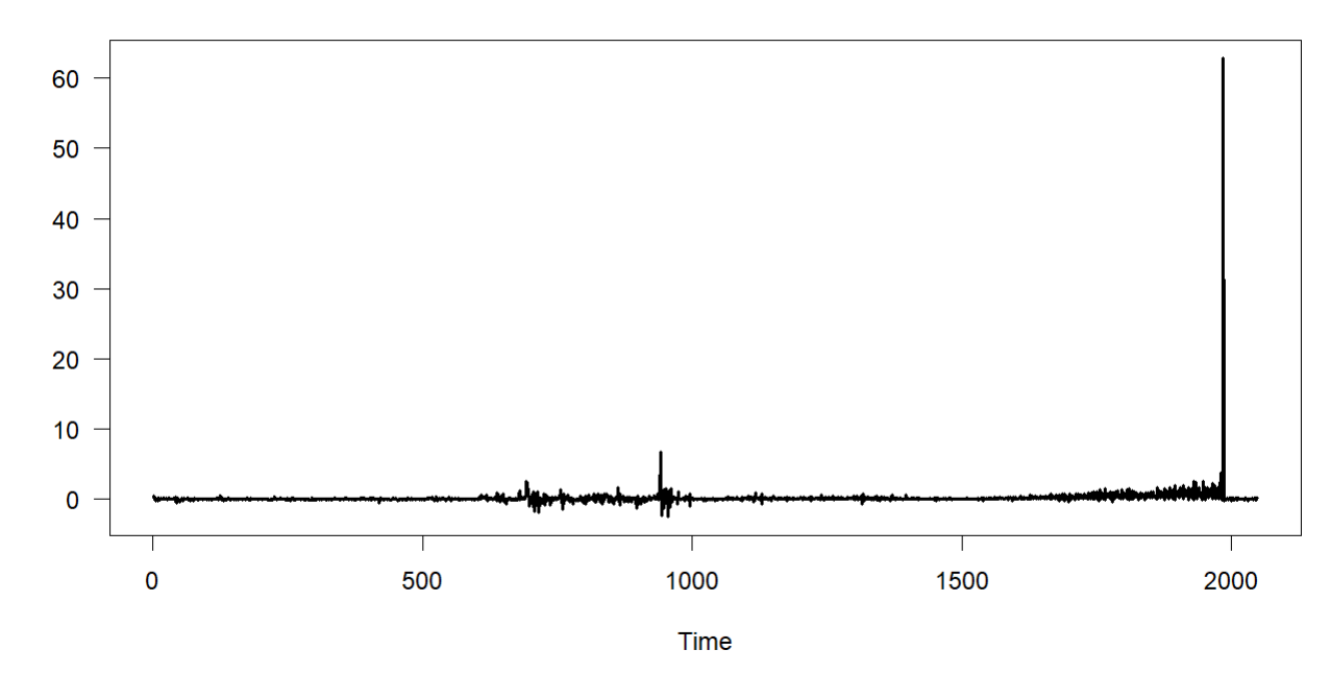}
		\caption{}
	\end{subfigure}
	\caption{(a) Plot of the daily minimum exchange rate of US Dollar versus Argentinian Peso from January 4, 2016, to November 9, 2023. (b) Time series plot of the first-order differenced series.}
	\label{daplots}
	\end{figure}

\begin{figure}
	\centering
	\begin{subfigure}[t]{0.48\textwidth}
		\centering
		\includegraphics[width=\textwidth]{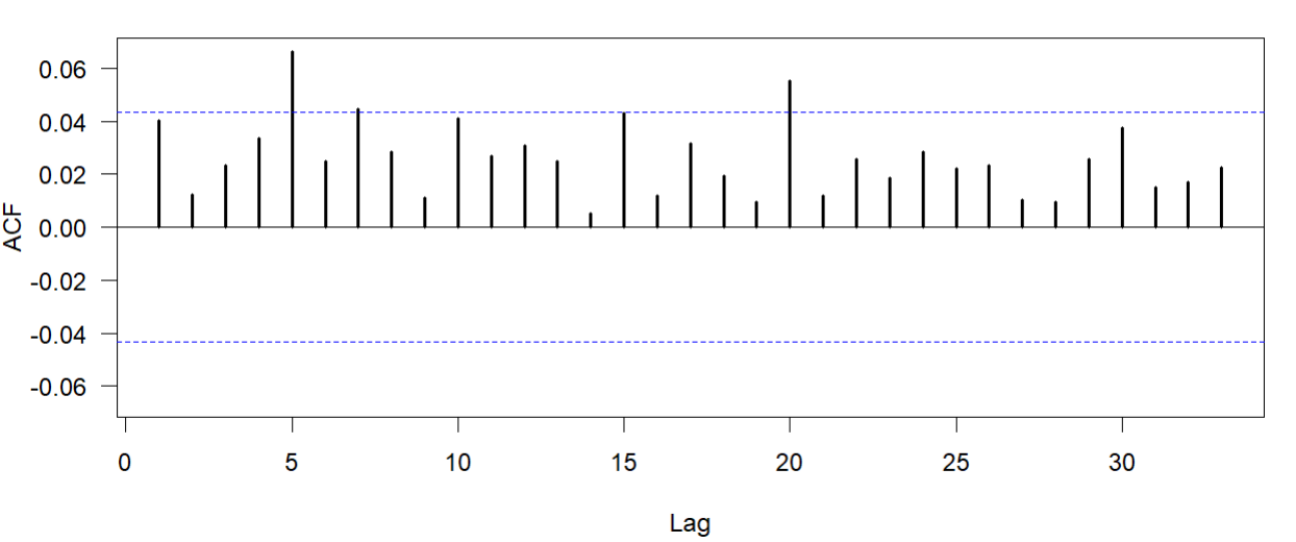}
		\caption{}
		\label{acfpltdata}
	\end{subfigure}
	\hfill
	\begin{subfigure}[t]{0.48\textwidth}
		\centering
		\includegraphics[width=\textwidth]{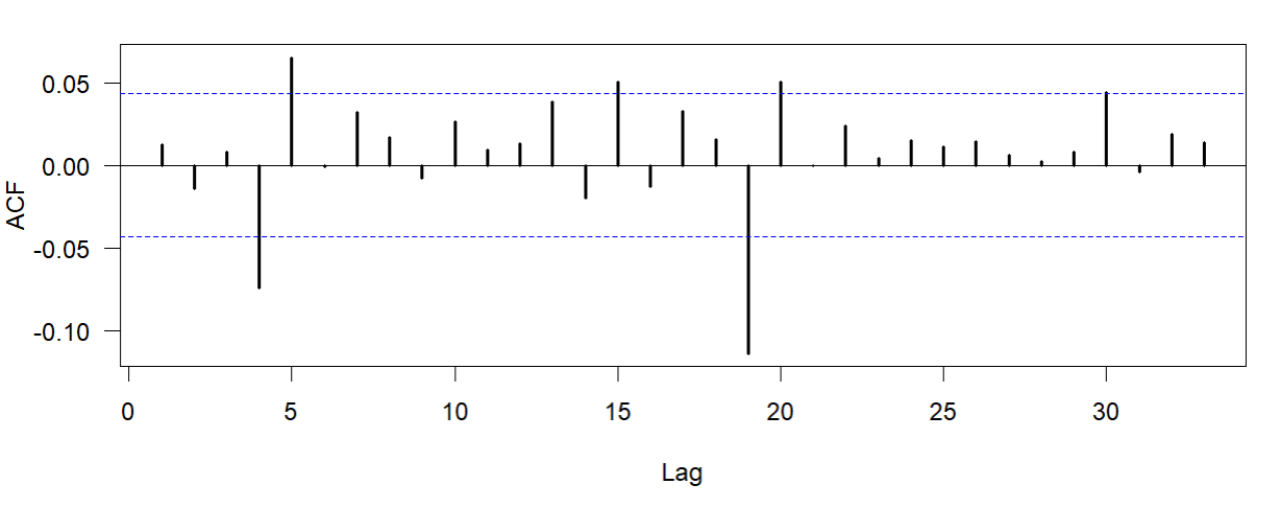}
		\caption{}
			\label{acfpltconst}
	\end{subfigure}	
	\vspace{0.3cm}	
	\begin{subfigure}[t]{0.49\textwidth}
		\centering
		\includegraphics[width=\textwidth]{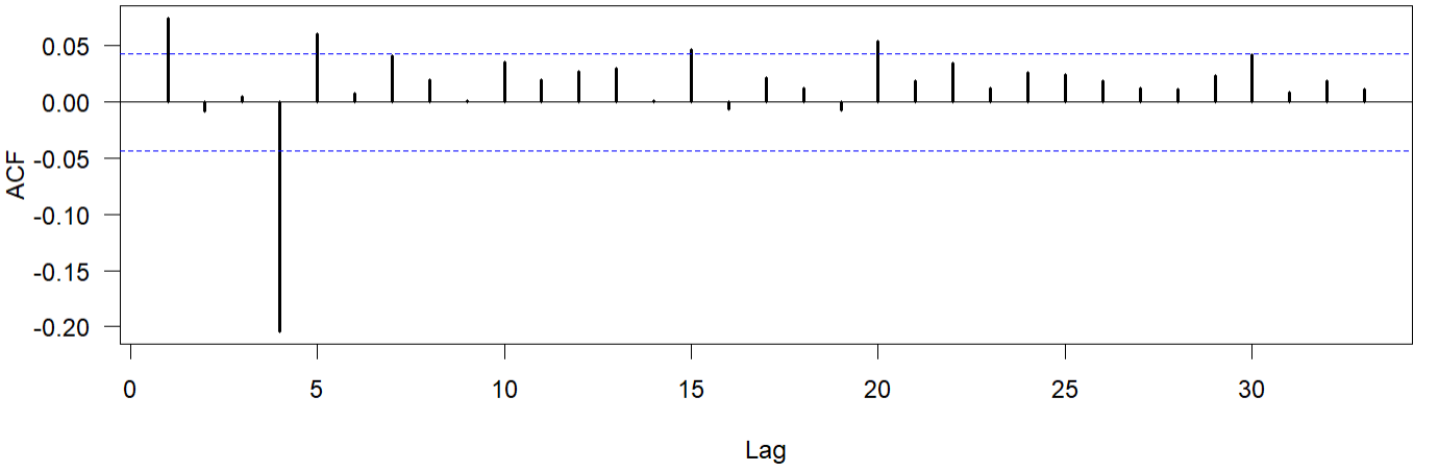}
		\caption{}
			\label{acfpltfou}
	\end{subfigure}
	\hfill
	\begin{subfigure}[t]{0.48\textwidth}
		\centering
		\includegraphics[width=\textwidth]{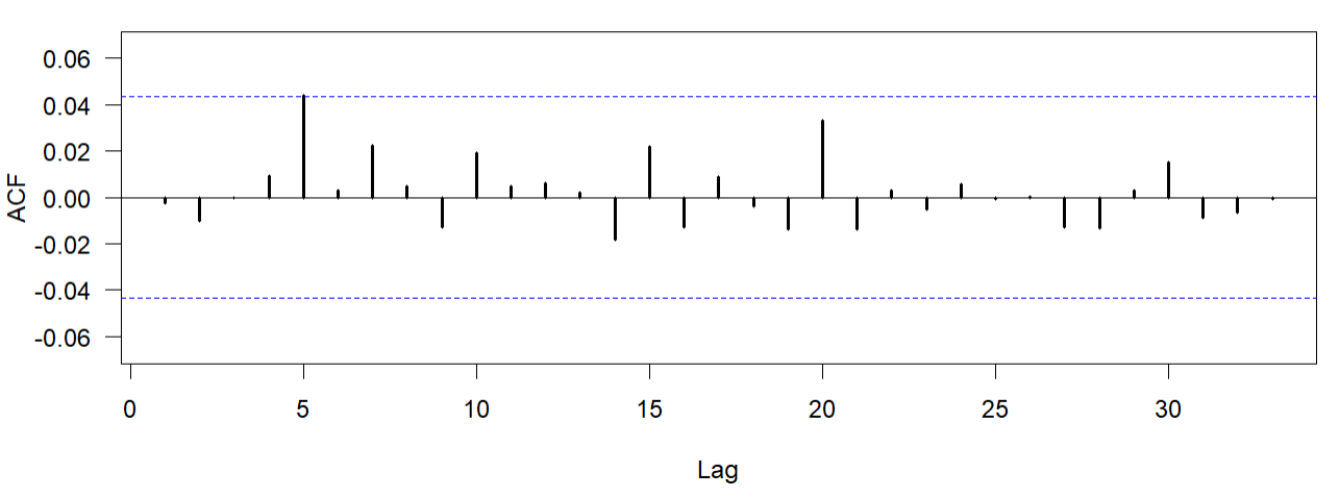}
		\caption{}
		\label{acfpltwav}
	\end{subfigure}
	\caption{(a) ACF plot of the differenced series. (b) ACF plot of the constant threshold model residuals. (c) ACF plot of the Fourier threshold model residuals. (d) ACF plot of the wavelet threshold model residuals.}
\end{figure}

In our study, we consider the daily minimum of the exchange rate of the US Dollar versus the Argentine Peso from January 4, 2016, to November 9, 2023, comprising 2048 observations. The data can be found in \textit{https://in.investing.com/currencies/usd-ars-historical-data}. For the analysis, the data were transformed by taking first-order difference. Figure \ref{daplots} presents the time series plots of the original series and its first difference. The ACF plot of the differenced series is shown in Figure \ref{acfpltdata}, indicating the presence of serial dependence. The Ljung–Box test applied at several lags provides additional support for this observation. For example, the test yields p-values of 0.0011 and 0.0028 at lags 20 and 30, respectively.  
A test for linearity based on \citeauthor{hansen1999testing} (\citeyear{hansen1999testing}) was carried out using the setarTest routine from the R package \texttt{tsDyn} (see \citeauthor{DiNarzo2009}, \citeyear{DiNarzo2009}). The obtained p-value (0.01) suggests the presence of a threshold structure in the series.

The estimated constant threshold model (see \eqref{constthreshmod}) is specified by
\begin{equation}\label{constthreshmodest}
	y(t) = \begin{cases}
		0.1078 + 0.1994y(t-1) + \epsilon(t), \,\,y(t-1) \leq 2.0623 \\
		7.4069 - 0.1346y(t-1) + \epsilon(t), \,\, y(t-1) > 2.0623
	\end{cases},\,\, t = 0, 1, \dotsc, 2047,
\end{equation}
where $\epsilon(t) \sim \text{iid}\,(0,1.8498)$. Figure \ref{acfpltconst} presents the autocorrelation function (ACF) of the residuals obtained from the constant threshold model. The Ljung-Box test produced p-values of $1.85 \times 10^{-7}$ and $5.92 \times 10^{-6}$ at lags 20 and 30, respectively. The results show significant serial correlation in the residuals, suggesting that the constant threshold model is not suitable for modeling the data’s dependence structure.

The parameters of the threshold model with a time-varying threshold, where the threshold is approximated using a Fourier series (see \eqref{tvarhreshmod} and \eqref{fouapp}), are estimated following the procedure proposed by \citeauthor{yang2021threshold} (\citeyear{yang2021threshold}). The frequency $k$ in \eqref{fouapp} is selected by implementing the optimization algorithm for $k=1,2,3,4,$ and 5 and choosing the value corresponding to the smallest SSR. The autoregressive parameters and the Fourier coefficients corresponding to the selected value of $k$ are then taken. The resulting fitted model is 
\begin{equation}
	y(t) = \begin{cases}
		0.1069 + 0.1950y(t-1) + \epsilon(t), \quad y(t-1) \leq \gamma \left(\frac{t}{T}\right) \\
		17.7802 - 0.2960y(t-1) + \epsilon(t), \quad y(t-1) > \gamma\left(\frac{t}{T}\right)
	\end{cases},
	\quad t = 0,1,\dotsc,2047,
\end{equation}
where 
$\epsilon(t) \sim \text{iid}\,(0,1.5240)$, and the estimated threshold function is
\begin{equation}
	\gamma \left(\frac{t}{T}\right) = 5.1946 - 0.0314 \sin\left(\frac{6\pi t}{T}\right)- 3.9465 \cos\left(\frac{6\pi t}{T}\right),
	\quad t = 0,\dotsc,2047.
\end{equation}
Figure \ref{acfpltfou} displays the autocorrelation function (ACF) of the residuals from the fitted model. The Ljung–Box test yields a p-value smaller than 
$2.2 \times 10^{-16}$ at lag 20 and a p-value of $4.44\times 10^{-16}$ at lag 30. These results show considerable serial dependence in the residuals, suggesting that the model fails to entirely capture the data's underlying dependence structure.

\begin{figure}
	\centering
	\includegraphics[scale = 0.4]{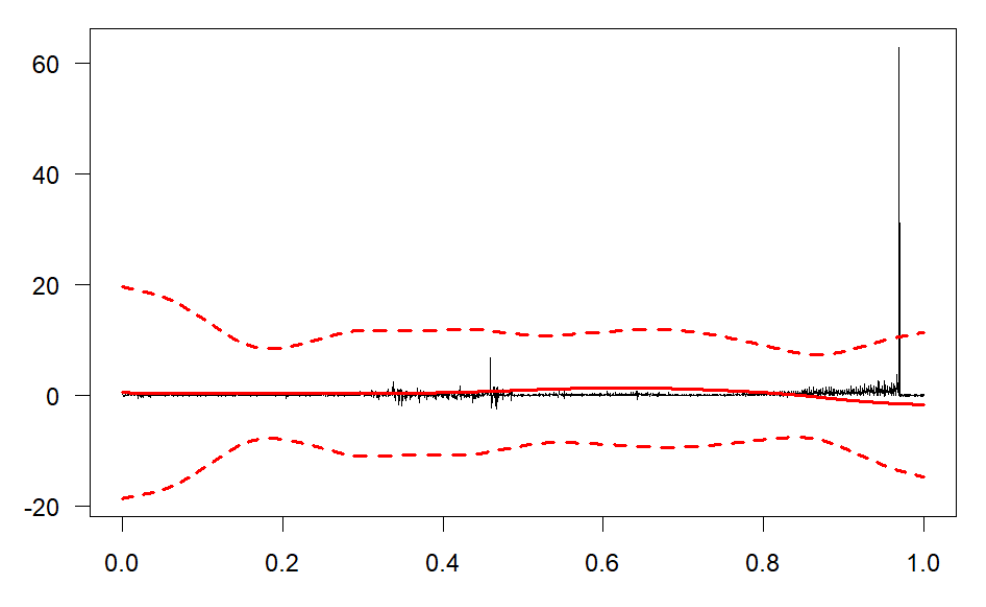}
	\caption{The black line represents the considered data, the red solid line shows the estimated wavelet-based threshold function and the red dotted lines present the bootstrap simulataneous bands via sup-t.}
	\label{daestwavthresh}
\end{figure}

The \texttt{wavethresh} package (see \citeauthor{wavethresh}, \citeyear{wavethresh}) provides the filter coefficients for the Daubechies extremal phase wavelet family with vanishing moments ranging from 1 to 10, as well as for the Daubechies least asymmetric wavelet family with vanishing moments from 4 to 10. The wavelet-based threshold models fitted using these wavelets produced comparable values of RMSE and MAE. For illustration, we report the results obtained using the D(5) wavelet. The estimated wavelet-based threshold model (see \eqref{wavethreshmod} and \eqref{wavserexp}) is given by
\begin{equation}
	y(t) = \begin{cases}
		0.0636 - 0.0710y(t-1) + \epsilon(t), \,\,y(t-1) \leq \gamma (\frac{t}{T}) \\
		0.5609 + 0.0158y(t-1) + \epsilon(t),\,\,y(t-1) > \gamma(\frac{t}{T})
	\end{cases},\,\, t = 0, \dotsc, 2047,
\end{equation}
where $\epsilon(t) \sim \text{iid}\,(0,2.0412)$. The optimal resolution level was found to be $J=3$, as it yielded the minimum RMSE among $J=2, 3, 4$, and 5. The bootstrap confidence intervals of $\phi_0^{(1)}, \phi_1^{(1)}, \phi_0^{(2)}, \phi_1^{(2)}$ and $\sigma^2$ obtained via the method in Section \ref{bootsec} with $B = 1000$ are $(0.0062, 0.4061), (-0.3720, 0.9184), (0.0917, 0.5927), (0.0084, 0.3439)$, and $(0.1298, 5.8621)$, respectively. Figure \ref{daestwavthresh} shows the estimated wavelet threshold curve (see \eqref{wavserexp}) and the bootstrap simultaneous sup-t bands superimposed on the data considered. 
The ACF plot of the residuals of this model is displayed in Figure \ref{acfpltwav}. The Ljung-Box test results at lags 20 and 30 are $0.9320$ and $0.9965$, respectively, indicating no significant serial correlation in the residuals. These results illustrate the proposed model's potential to capture the data's underlying dependence structure.

\section{Conclusion}\label{sec13}
Threshold autoregressive models have been used when a single linear model cannot capture the different regimes in the process. A constant threshold is often inadequate because a variable's dynamics are determined by other time-varying variables. A threshold model with a threshold that varies over time, estimated via a wavelet series expansion, is proposed and is especially effective when the threshold parameter undergoes sudden changes. Simulation and real data analyses demonstrate the effectiveness of the proposed method, and the approach shows promise for extension to other classes of threshold models.\\

\hfill\\
\textbf{Acknowledgements}\\
The authors acknowledge access to the High Performance Computing (HPC) facility at CUSAT as part of the RUSA 2.0 Project T3A. Rhea Davis acknowledges financial support from the University Grants Commission (UGC) of India through the Savitribai Jyotirao Phule Fellowship for Single Girl Child (SJSGC) scheme.

\bibliography{sn_article}


\begin{thebibliography}{30}
\ifx \bisbn   \undefined \def \bisbn  #1{ISBN #1}\fi
\ifx \binits  \undefined \def \binits#1{#1}\fi
\ifx \bauthor  \undefined \def \bauthor#1{#1}\fi
\ifx \batitle  \undefined \def \batitle#1{#1}\fi
\ifx \bjtitle  \undefined \def \bjtitle#1{#1}\fi
\ifx \bvolume  \undefined \def \bvolume#1{\textbf{#1}}\fi
\ifx \byear  \undefined \def \byear#1{#1}\fi
\ifx \bissue  \undefined \def \bissue#1{#1}\fi
\ifx \bfpage  \undefined \def \bfpage#1{#1}\fi
\ifx \blpage  \undefined \def \blpage #1{#1}\fi
\ifx \burl  \undefined \def \burl#1{\textsf{#1}}\fi
\ifx \doiurl  \undefined \def \doiurl#1{\url{https://doi.org/#1}}\fi
\ifx \betal  \undefined \def \betal{\textit{et al.}}\fi
\ifx \binstitute  \undefined \def \binstitute#1{#1}\fi
\ifx \binstitutionaled  \undefined \def \binstitutionaled#1{#1}\fi
\ifx \bctitle  \undefined \def \bctitle#1{#1}\fi
\ifx \beditor  \undefined \def \beditor#1{#1}\fi
\ifx \bpublisher  \undefined \def \bpublisher#1{#1}\fi
\ifx \bbtitle  \undefined \def \bbtitle#1{#1}\fi
\ifx \bedition  \undefined \def \bedition#1{#1}\fi
\ifx \bseriesno  \undefined \def \bseriesno#1{#1}\fi
\ifx \blocation  \undefined \def \blocation#1{#1}\fi
\ifx \bsertitle  \undefined \def \bsertitle#1{#1}\fi
\ifx \bsnm \undefined \def \bsnm#1{#1}\fi
\ifx \bsuffix \undefined \def \bsuffix#1{#1}\fi
\ifx \bparticle \undefined \def \bparticle#1{#1}\fi
\ifx \barticle \undefined \def \barticle#1{#1}\fi
\bibcommenthead
\ifx \bconfdate \undefined \def \bconfdate #1{#1}\fi
\ifx \botherref \undefined \def \botherref #1{#1}\fi
\ifx \url \undefined \def \url#1{\textsf{#1}}\fi
\ifx \bchapter \undefined \def \bchapter#1{#1}\fi
\ifx \bbook \undefined \def \bbook#1{#1}\fi
\ifx \bcomment \undefined \def \bcomment#1{#1}\fi
\ifx \oauthor \undefined \def \oauthor#1{#1}\fi
\ifx \citeauthoryear \undefined \def \citeauthoryear#1{#1}\fi
\ifx \endbibitem  \undefined \def \endbibitem {}\fi
\ifx \bconflocation  \undefined \def \bconflocation#1{#1}\fi
\ifx \arxivurl  \undefined \def \arxivurl#1{\textsf{#1}}\fi
\csname PreBibitemsHook\endcsname

\bibitem[\protect\citeauthoryear{Tsay}{2005}]{tsay2005analysis}
\begin{bbook}
\bauthor{\bsnm{Tsay}, \binits{R.S.}}:
\bbtitle{Analysis of Financial Time Series}.
\bpublisher{John Wiley \& Sons},
\blocation{Hoboken, NJ}
(\byear{2005})
\end{bbook}
\endbibitem

\bibitem[\protect\citeauthoryear{Tong}{1978}]{tong1978threshold}
\begin{botherref}
\oauthor{\bsnm{Tong}, \binits{H.}}:
On a threshold model.
Pattern Recognition and Signal Processing,
575--586
(1978)
\end{botherref}
\endbibitem

\bibitem[\protect\citeauthoryear{Yadav et~al.}{1994}]{yadav1994threshold}
\begin{barticle}
\bauthor{\bsnm{Yadav}, \binits{P.K.}},
\bauthor{\bsnm{Pope}, \binits{P.F.}},
\bauthor{\bsnm{Paudyal}, \binits{K.}}:
\batitle{Threshold autoregressive modeling in finance: The price differences of
  equivalent assets}.
\bjtitle{Mathematical Finance}
\bvolume{4}(\bissue{2}),
\bfpage{205}--\blpage{221}
(\byear{1994})
\end{barticle}
\endbibitem

\bibitem[\protect\citeauthoryear{Petruccelli and
  Woolford}{1984}]{petruccelli1984threshold}
\begin{barticle}
\bauthor{\bsnm{Petruccelli}, \binits{J.D.}},
\bauthor{\bsnm{Woolford}, \binits{S.W.}}:
\batitle{A threshold {AR} (1) model}.
\bjtitle{Journal of Applied Probability}
\bvolume{21}(\bissue{2}),
\bfpage{270}--\blpage{286}
(\byear{1984})
\end{barticle}
\endbibitem

\bibitem[\protect\citeauthoryear{Tsay}{1989}]{tsay1989testing}
\begin{barticle}
\bauthor{\bsnm{Tsay}, \binits{R.S.}}:
\batitle{Testing and modeling threshold autoregressive processes}.
\bjtitle{Journal of the American Statistical Association}
\bvolume{84}(\bissue{405}),
\bfpage{231}--\blpage{240}
(\byear{1989})
\end{barticle}
\endbibitem

\bibitem[\protect\citeauthoryear{Sirikanchanarak
  et~al.}{2016}]{sirikanchanarak2016time}
\begin{barticle}
\bauthor{\bsnm{Sirikanchanarak}, \binits{D.}},
\bauthor{\bsnm{Yamaka}, \binits{W.}},
\bauthor{\bsnm{Khiewgamdee}, \binits{C.}},
\bauthor{\bsnm{Sriboonchitta}, \binits{S.}}:
\batitle{Time-varying threshold regression model using the kalman filter
  method}.
\bjtitle{Thai Journal of Mathematics}
\bvolume{74},
\bfpage{133}--\blpage{148}
(\byear{2016})
\end{barticle}
\endbibitem

\bibitem[\protect\citeauthoryear{Montgomery
  et~al.}{1998}]{montgomery1998forecasting}
\begin{barticle}
\bauthor{\bsnm{Montgomery}, \binits{A.L.}},
\bauthor{\bsnm{Zarnowitz}, \binits{V.}},
\bauthor{\bsnm{Tsay}, \binits{R.S.}},
\bauthor{\bsnm{Tiao}, \binits{G.C.}}:
\batitle{Forecasting the {U.S.} unemployment rate}.
\bjtitle{Journal of the American Statistical Association}
\bvolume{93}(\bissue{442}),
\bfpage{478}--\blpage{493}
(\byear{1998})
\end{barticle}
\endbibitem

\bibitem[\protect\citeauthoryear{Li and Ling}{2012}]{li2012least}
\begin{barticle}
\bauthor{\bsnm{Li}, \binits{D.}},
\bauthor{\bsnm{Ling}, \binits{S.}}:
\batitle{On the least squares estimation of multiple-regime threshold
  autoregressive models}.
\bjtitle{Journal of Econometrics}
\bvolume{167}(\bissue{1}),
\bfpage{240}--\blpage{253}
(\byear{2012})
\end{barticle}
\endbibitem

\bibitem[\protect\citeauthoryear{Narayan}{2006}]{narayan2006behaviour}
\begin{barticle}
\bauthor{\bsnm{Narayan}, \binits{P.K.}}:
\batitle{The behaviour of {US} stock prices: Evidence from a threshold
  autoregressive model}.
\bjtitle{Mathematics and Computers in Simulation}
\bvolume{71}(\bissue{2}),
\bfpage{103}--\blpage{108}
(\byear{2006})
\end{barticle}
\endbibitem

\bibitem[\protect\citeauthoryear{Clements and Smith}{1999}]{clements1999monte}
\begin{barticle}
\bauthor{\bsnm{Clements}, \binits{M.P.}},
\bauthor{\bsnm{Smith}, \binits{J.}}:
\batitle{A monte carlo study of the forecasting performance of empirical setar
  models}.
\bjtitle{Journal of Applied Econometrics}
\bvolume{14}(\bissue{2}),
\bfpage{123}--\blpage{141}
(\byear{1999})
\end{barticle}
\endbibitem

\bibitem[\protect\citeauthoryear{Huang}{1997}]{huang1997short}
\begin{barticle}
\bauthor{\bsnm{Huang}, \binits{S.}}:
\batitle{Short-term load forecasting using threshold autoregressive models}.
\bjtitle{IEE Proceedings-Generation, Transmission and Distribution}
\bvolume{144}(\bissue{5}),
\bfpage{477}--\blpage{481}
(\byear{1997})
\end{barticle}
\endbibitem

\bibitem[\protect\citeauthoryear{Watier and
  Richardson}{1995}]{watier1995modelling}
\begin{barticle}
\bauthor{\bsnm{Watier}, \binits{L.}},
\bauthor{\bsnm{Richardson}, \binits{S.}}:
\batitle{Modelling of an epidemiological time series by a threshold
  autoregressive model}.
\bjtitle{Journal of the Royal Statistical Society: Series D (The Statistician)}
\bvolume{44}(\bissue{3}),
\bfpage{353}--\blpage{364}
(\byear{1995})
\end{barticle}
\endbibitem

\bibitem[\protect\citeauthoryear{Hansen}{1997}]{hansen1997inference}
\begin{barticle}
\bauthor{\bsnm{Hansen}, \binits{B.E.}}:
\batitle{{I}nference in {TAR} models}.
\bjtitle{Studies in Nonlinear Dynamics and Econometrics}
\bvolume{1},
\bfpage{119}--\blpage{131}
(\byear{1997})
\end{barticle}
\endbibitem

\bibitem[\protect\citeauthoryear{Hansen}{2011}]{hansen2011threshold}
\begin{barticle}
\bauthor{\bsnm{Hansen}, \binits{B.E.}}:
\batitle{Threshold autoregression in economics}.
\bjtitle{Statistics and its Interface}
\bvolume{4}(\bissue{2}),
\bfpage{123}--\blpage{127}
(\byear{2011})
\end{barticle}
\endbibitem

\bibitem[\protect\citeauthoryear{Zhu and Chen}{2017}]{zhu2017asymmetry}
\begin{barticle}
\bauthor{\bsnm{Zhu}, \binits{Y.}},
\bauthor{\bsnm{Chen}, \binits{H.}}:
\batitle{The asymmetry of {US} monetary policy: Evidence from a threshold
  {T}aylor rule with time-varying threshold values}.
\bjtitle{Physica A: Statistical Mechanics and its Applications}
\bvolume{473},
\bfpage{522}--\blpage{535}
(\byear{2017})
\end{barticle}
\endbibitem

\bibitem[\protect\citeauthoryear{Yang et~al.}{2021}]{yang2021threshold}
\begin{barticle}
\bauthor{\bsnm{Yang}, \binits{L.}},
\bauthor{\bsnm{Lee}, \binits{C.}},
\bauthor{\bsnm{Chen}, \binits{I.-P.}}:
\batitle{Threshold model with a time-varying threshold based on {F}ourier
  approximation}.
\bjtitle{Journal of Time Series Analysis}
\bvolume{42}(\bissue{4}),
\bfpage{406}--\blpage{430}
(\byear{2021})
\end{barticle}
\endbibitem

\bibitem[\protect\citeauthoryear{Dueker et~al.}{2013}]{dueker2013state}
\begin{barticle}
\bauthor{\bsnm{Dueker}, \binits{M.J.}},
\bauthor{\bsnm{Psaradakis}, \binits{Z.}},
\bauthor{\bsnm{Sola}, \binits{M.}},
\bauthor{\bsnm{Spagnolo}, \binits{F.}}:
\batitle{State-dependent threshold smooth transition autoregressive models}.
\bjtitle{Oxford Bulletin of Economics and Statistics}
\bvolume{75}(\bissue{6}),
\bfpage{835}--\blpage{854}
(\byear{2013})
\end{barticle}
\endbibitem

\bibitem[\protect\citeauthoryear{Bessec}{2003}]{bessec2003asymmetric}
\begin{barticle}
\bauthor{\bsnm{Bessec}, \binits{M.}}:
\batitle{The asymmetric exchange rate dynamics in the {EMS}: a time-varying
  threshold test}.
\bjtitle{European Review of Economics and Finance}
\bvolume{2}(\bissue{2}),
\bfpage{3}--\blpage{40}
(\byear{2003})
\end{barticle}
\endbibitem

\bibitem[\protect\citeauthoryear{Yu and Fan}{2021}]{yu2021threshold}
\begin{barticle}
\bauthor{\bsnm{Yu}, \binits{P.}},
\bauthor{\bsnm{Fan}, \binits{X.}}:
\batitle{Threshold regression with a threshold boundary}.
\bjtitle{Journal of Business \& Economic Statistics}
\bvolume{39}(\bissue{4}),
\bfpage{953}--\blpage{971}
(\byear{2021})
\end{barticle}
\endbibitem

\bibitem[\protect\citeauthoryear{Daubechies}{1992}]{daubechies1992ten}
\begin{bbook}
\bauthor{\bsnm{Daubechies}, \binits{I.}}:
\bbtitle{Ten Lectures on Wavelets}.
\bpublisher{SIAM},
\blocation{Philadelphia, PA}
(\byear{1992})
\end{bbook}
\endbibitem

\bibitem[\protect\citeauthoryear{Bak et~al.}{2025}]{bak2025adaptive}
\begin{botherref}
\oauthor{\bsnm{Bak}, \binits{K.-Y.}},
\oauthor{\bsnm{Lee}, \binits{E.-J.}},
\oauthor{\bsnm{Jhong}, \binits{J.-H.}}:
Adaptive log-wavelet density estimation with resolution identification.
Journal of the Korean Statistical Society,
1--25
(2025)
\end{botherref}
\endbibitem

\bibitem[\protect\citeauthoryear{Chen et~al.}{2025}]{chen2025time}
\begin{barticle}
\bauthor{\bsnm{Chen}, \binits{Y.}},
\bauthor{\bsnm{Morettin}, \binits{P.A.}},
\bauthor{\bsnm{Chiann}, \binits{C.}}:
\batitle{Time-varying spatio-temporal models by wavelets}.
\bjtitle{Journal of Statistical Computation and Simulation}
\bvolume{95}(\bissue{16}),
\bfpage{3442}--\blpage{3468}
(\byear{2025})
\end{barticle}
\endbibitem

\bibitem[\protect\citeauthoryear{Morettin et~al.}{2017}]{morettin2017wavelets}
\begin{bbook}
\bauthor{\bsnm{Morettin}, \binits{P.A.}},
\bauthor{\bsnm{Pinheiro}, \binits{A.}},
\bauthor{\bsnm{Vidakovic}, \binits{B.}}:
\bbtitle{Wavelets in Functional Data Analysis}.
\bpublisher{Springer},
\blocation{Switzerland}
(\byear{2017})
\end{bbook}
\endbibitem

\bibitem[\protect\citeauthoryear{Haar}{1910}]{haar1910theorie}
\begin{barticle}
\bauthor{\bsnm{Haar}, \binits{A.}}:
\batitle{Zur {Theorie} der orthogonalen {Funktionen-Systeme}}.
\bjtitle{Math. Ann.}
\bvolume{69},
\bfpage{331}--\blpage{371}
(\byear{1910})
\end{barticle}
\endbibitem

\bibitem[\protect\citeauthoryear{H{\"a}rdle et~al.}{2012}]{hardle2012wavelets}
\begin{bbook}
\bauthor{\bsnm{H{\"a}rdle}, \binits{W.}},
\bauthor{\bsnm{Kerkyacharian}, \binits{G.}},
\bauthor{\bsnm{Picard}, \binits{D.}},
\bauthor{\bsnm{Tsybakov}, \binits{A.}}:
\bbtitle{Wavelets, Approximation, and Statistical Applications}
vol. \bseriesno{129}.
\bpublisher{Springer},
\blocation{New York}
(\byear{2012})
\end{bbook}
\endbibitem

\bibitem[\protect\citeauthoryear{Mullen et~al.}{2011}]{deoptimpack}
\begin{barticle}
\bauthor{\bsnm{Mullen}, \binits{K.}},
\bauthor{\bsnm{Ardia}, \binits{D.}},
\bauthor{\bsnm{Gil}, \binits{D.}},
\bauthor{\bsnm{Windover}, \binits{D.}},
\bauthor{\bsnm{Cline}, \binits{J.}}:
\batitle{{DEoptim}: An {R} package for global optimization by {D}ifferential
  {E}volution}.
\bjtitle{Journal of Statistical Software}
\bvolume{40}(\bissue{6}),
\bfpage{1}--\blpage{26}
(\byear{2011})
\end{barticle}
\endbibitem

\bibitem[\protect\citeauthoryear{Montiel~Olea and
  Plagborg-M{\o}ller}{2019}]{montiel2019simultaneous}
\begin{barticle}
\bauthor{\bsnm{Montiel~Olea}, \binits{J.L.}},
\bauthor{\bsnm{Plagborg-M{\o}ller}, \binits{M.}}:
\batitle{Simultaneous confidence bands: Theory, implementation, and an
  application to svars}.
\bjtitle{Journal of Applied Econometrics}
\bvolume{34}(\bissue{1}),
\bfpage{1}--\blpage{17}
(\byear{2019})
\end{barticle}
\endbibitem

\bibitem[\protect\citeauthoryear{Hansen}{1999}]{hansen1999testing}
\begin{barticle}
\bauthor{\bsnm{Hansen}, \binits{B.}}:
\batitle{Testing for linearity}.
\bjtitle{Journal of Economic Surveys}
\bvolume{13}(\bissue{5}),
\bfpage{551}--\blpage{576}
(\byear{1999})
\end{barticle}
\endbibitem

\bibitem[\protect\citeauthoryear{{Fabio Di Narzo} et~al.}{2009}]{DiNarzo2009}
\begin{botherref}
\oauthor{\bsnm{{Fabio Di Narzo}}, \binits{A.}},
\oauthor{\bsnm{Aznarte}, \binits{J.L.}},
\oauthor{\bsnm{Stigler}, \binits{M.}}:
tsDyn: Time Series Analysis Based on Dynamical Systems Theory.
(2009).
R package version 0.7
\end{botherref}
\endbibitem

\bibitem[\protect\citeauthoryear{Nason}{2024}]{wavethresh}
\begin{botherref}
\oauthor{\bsnm{Nason}, \binits{G.}}:
Wavethresh: Wavelets Statistics and Transforms.
(2024).
R package version 4.7.3.
\url{https://CRAN.R-project.org/package=wavethresh}
\end{botherref}
\endbibitem

\end{thebibliography}
\end{document}